\documentclass[12pt]{article}
\usepackage{authblk}
\usepackage{amsmath}
\usepackage{amssymb}
\usepackage[hmargin=20mm, vmargin=20mm]{geometry}
\usepackage{graphicx}
\usepackage{subfig}
\usepackage[font=scriptsize]{caption}

\begin{document}
\title{Thermodynamics of a static electric-magnetic black hole in Einstein-Born-Infeld-AdS theory with different horizon geometries}
\author{M. B. Tataryn\footnote{E-mail: misha.physics@gmail.com}, M. M. Stetsko\footnote{E-mail: mstetsko@gmail.com}}
\affil{\small Department for Theoretical Physics,\\Ivan Franko National University of Lviv,\\12 Drahomanov Str., Lviv, 79005, Ukraine}
\date{}
\maketitle
\begin{abstract}
We consider black hole solutions with electric and magnetic sources in the four-dimensional Einstein-Born-Infeld-AdS theory with spherical, planar and hyperbolic horizon geometries. Exact analytical solutions for the metric function, electric and magnetic fields were obtained and they recover the RN-AdS black hole in the limit $\beta\to+\infty$ for spherical horizon in the absence of the magnetic charge. Expressions for temperature, electric and magnetic potential were obtained and they satisfy the first law of the extended black hole thermodynamics, where a negative cosmological constant is associated with thermodynamic pressure. Also, the Born-Infeld vacuum polarization term $Bd\beta$ was included into the first law in order to satisfy the Smarr relation. Critical behavior of the black hole was examined and condition on electric and magnetic charges were obtained when phase transition appears. Also, the critical ratio and capacity at constant pressure were calculated. Electric and magnetic charges enter into the metric function and thermodynamic quantities symmetrically and thus the presence of the magnetic charge does not bring very significant new features. Finally, we examine the Joule-Thomson expansion if the black hole mass is fixed. The inversion and isenthalpic curves were plotted and the cooling and heating regions were demonstrated. These results recover the Joule-Thomson expansion recently considered for the RN-AdS black hole in the corresponding limit.

\end{abstract}
\section{Introduction}
The Reissner-Nordstr{\"o}m-AdS solution in the anti-de Sitter spacetime (RN-AdS black hole) is the well-known solution of the four-dimensional, non-rotating, electrically charged black hole with negative cosmological constant. This solution is obtained in framework of General Relativity with Maxwell electrodynamics. The thermodynamic behavior of this black hole is examined by many researchers in many aspects today. One of the useful ideas consists in considering of the cosmological constant as one of the variable thermodynamic quantities, namely the thermodynamic pressure \cite{kas09,dol11,dol11_2,sek06,cvet11,kub17}. This allows to consider an additional term in the first law of black hole thermodynamics ($VdP$ term), where the pressure enters together with its conjugate quantity (that is thermodynamic volume), and in such extended phase space the mass of the black hole is identified with enthalpy. Such approach allows to set some analogy between the charged RN-AdS black hole \cite{kub12} and the Van der Waals gas, particularly the universal critical ratio $P_cv_c/T_c=3/8$ is the same for the RN-AdS and the Van der Waals systems. Phase transitions of rotating black holes and black rings were considered in \cite{altam14}.

Authors of \cite{okcu17} have considered the Joule-Thomson expansion of the charged RN-AdS black hole in extended phase space and obtained corresponding inversion and isoenthalpic curves. The Joule-Thomson expansion of $d-$dimensional charged AdS black holes is reviewed in \cite{mo18}, whereas the Joule-Thomson expansion of black holes with momentum relaxation was considered in \cite{cist19}. The Joule-Thomson expansion of the Born-Infeld AdS black hole was investigated in a recent paper \cite{bi21}.

On the other hand today various nonlinear generalizations of the Maxwell electrodynamics are considered actively. Among them we point out the Power-Maxwell theory, where electromagnetic Lagrangian is chosen to be $(-F_{\mu\nu}F^{\mu\nu})^s$ with rational power $s$, in this context there are three-dimensional solution \cite{mazh14}, higher-dimensional black holes with conformall source ($s=d/4$) \cite{has07}, and with arbitrary power \cite{has08}, slowly rotating black holes \cite{hen10,pan19,tat20}, thermodynamic stability \cite{gon09} and many others. Another nonlinear electrodynamics realization is original Born-Infeld theory \cite{born34}, which gives finite self-energy of a point-like electric charge. Electrically charged AdS black hole solution in Einstein-Born-Infeld gravity can be found in \cite{fer03}. A phase transition of the electrically charged Born-Infeld AdS black hole, its isothermal compressibility, and critical exponents were examined in four \cite{ban12}, and higher \cite{ban12_2} dimensional cases, respectively. There are many papers about various aspects of black hole solution and its thermodynamic behavior in the Born-Infeld theory \cite{li16,myun08,dey04,cai04,mis08,wei10,gunas12,zou14}, quasinormal modes \cite{bret17}. Authors of \cite{li16} consider dyonic black holes, their thermodynamic quantities and the first law. Other Born-Infeld type fields as the logarithmic and exponential ones were considered, for example, in \cite{hen_all_14,shey14,hen16,tat19} and the Born-Infeld inspired modifications of gravity in \cite{jimen18}. Some other realizations of nonlinear electrodynamics see, for example, in \cite{gaet17,krugl15}.

One more possibility which generalizes the Maxwell electrodynamics is consideration of magnetic charges.
Due to electric-magnetic duality in String Theory black holes with magnetic (and electric) charges have gained some attention and several new black hole solutions were derived in different settings of low energy limit of the String Theory \cite{shap91,sen93} and supergravity \cite{cham20}. Deep interest to dyonic (electric-magnetic) black holes is also motivated by further development of AdS/CFT correspondence, namely its application to various problems in Condensed Matter Physics, in particular we point out successful application of a dyonic black hole solution to obtain Hall conductivity \cite{hart07} or for description of the Nernst effect near superfluid-insulator quantum phase transition \cite{hart07_2}. In the context of AdS/CFT correspondence dyonic black holes were utilized to describe holographic superconductors \cite{alba08}. A non-Abelian Einstein-Born-Infeld black holes with purely magnetic gauge field were examined in \cite{wirs01}. Thermodynamic of the dyonic black holes also has become the subject of research in different directions. Thermodynamic properties of the AdS dyonic black holes were studied in \cite{lu13}. Equation of state, critical exponents, thermodynamic stability were investigated in \cite{dut13}. Also in that paper magnetic properties of boundary CFT were examined. Author of \cite{eiroa06} has studied gravitational lensing by four-dimensional Einstein-Born-Infeld black holes with electric and magnetic charges.

These various applications of dyonic black hole solutions show their importance and might be a good ground for their further investigation in various setting, in particular if the gauge field is supposed to be nonlinear. For example, in paper \cite{gib95} electric-magnetic duality rotations in non-linear electrodynamics was considered, and as an example applied to the Born-Infeld Lagrangian. Particularly, in that paper it was noticed, that the gravitational field with electric $q_e$ and magnetic $q_m$ sources depends only on the combination $\sqrt{q_e^2+q_m^2}$. In \cite{stef07} authors consider scalar-tensor magnetically charged black holes coupled to the Born-Infeld electrodynamics. Thermodynamics of the Born-Infeld black hole with electric and magnetic sources without of the cosmological constant has been studied in \cite{chem08}. Author of \cite{krug17,krug19} considers a magnetic black hole in the Born-Infeld type electrodynamics. Particularly, in \cite{krug19} a Born-Infeld type Lagrangian is considered where original Born-Infeld power $1/2$ is replaced to be $3/4$, and also only one electromagnetic invariant $F$ is taken. Dyonic black hole with the Born-Infeld field in Horndeski gravity was considered in a recent paper \cite{meng21}. This theory brings its distinctive features compared to the Einstein-Born-Infeld theory.

In this paper we consider a four-dimensional electrically and magnetically charged black hole with original Born-Infeld electromagnetic field in the anti-de Sitter space. We examine thermodynamics of this black hole, mainly its first law, phase transition and Joule-Thomson expansion in extended phase space. We study their dependence on the Born-Infeld nonlinearity parameter, electric and magnetic charges, and also compare obtained results with the RN-AdS case. We consider conditions on the Born-Infeld parameter, electric and magnetic charges for the existence of the black hole phase transition and Joule-Thomson expansion. Besides, we consider, so-called, topological solutions of the four-dimensional black hole with electric and magnetic charges in the Born-Infeld theory with planar and hyperbolic horizon geometries.

The paper is organized as follows. In the next section we give a four-dimensional black hole solution with electric and magnetic charges in the Born-Infeld electrodynamics for spherical, planar and hyperbolic horizon structures. In section \ref{thermodynamics} we consider thermodynamic quantities, first law, phase transition, and the Joule-Thomson expansion in extended phase space. In section \ref{conclusions} we briefly conclude our work.
\section{Black hole solution}\label{solution}
Hereinafter we use geometric units in which $G_N=c=\hbar=k_B=1$. We are going to study thermodynamic properties of a four-dimensional black hole with negative cosmological constant in Einstein gravity with the Born-Infeld electromagnetic field. The corresponding action is
\begin{equation}\label{action}
S=\frac{1}{16\pi}\int d^4x\sqrt{-g}(R-2\Lambda+L_{BI}),
\end{equation}
where $g=\det g_{\mu\nu}$, $g_{\mu\nu}$ is the metric tensor, $R$ is the scalar curvature, $\Lambda=-3/l^2$ is the cosmological constant, $L_{BI}$ is the Born-Infeld Lagrangian \cite{born34}
\begin{equation}
L_{BI}=4\beta^2\left(1-\sqrt{1+F-G^2}\right),
\end{equation}
where $\beta$ is the nonlinearity parameter, $F$ and $G$ are field invariants
\begin{equation}
F=\frac{1}{2\beta^2}F_{\alpha\beta}F^{\alpha\beta}=\frac{1}{\beta^2}(B^2-E^2),\qquad G=\frac{1}{4\beta^2}F_{\alpha\beta}F^{\ast\alpha\beta}=\frac{1}{\beta^2}(E\cdot B),
\end{equation}
where $F^{\ast\alpha\beta}=j^{\alpha\beta\gamma\delta}F_{\gamma\delta}$ is the dual tensor to $F_{\alpha\beta}$, $F_{\alpha\beta}=\partial_\alpha A_\beta-\partial_\beta A_\alpha$ is the electromagnetic tensor, $A_\alpha$ is the electromagnetic potential, $j^{\alpha\beta\gamma\delta}=(2\sqrt{-g})^{-1}\varepsilon^{\alpha\beta\gamma\delta}$, $\varepsilon^{\alpha\beta\gamma\delta}$ is the unit Levi-Civita tensor, $\varepsilon^{0123}=1$. Parameter $\beta>0$ by $\beta\to+\infty$ leads to the Maxwell Lagrangian, $L_{BI}\to-F_{\alpha\beta}F^{\alpha\beta}$ and the action (\ref{action}) goes to the RN-AdS ones. We note, that we use original form of the Born-Infeld Lagrangian with two field invariants $F$ and $G$.

The field equations read
\begin{equation}\label{grav}
R_{\mu\nu}-\frac{1}{2}g_{\mu\nu}R+g_{\mu\nu}\Lambda=\frac{1}{2}g_{\mu\nu}L_{BI}+\frac{2(F_{\mu\rho}F_\nu^{\ \rho}-g_{\mu\nu}\beta^2G^2)}{\sqrt{1+F-G^2}},
\end{equation}
\begin{equation}\label{em}
\partial_\mu\left[\frac{\sqrt{-g}(F^{\mu\nu}-GF^{\ast\mu\nu})}{\sqrt{1+F-G^2}}\right]=0,
\end{equation}
where $R_{\mu\nu}$ is the Ricci tensor. We are studying a static four-dimensional solution with spherical ($k=1$), planar (toroidal) ($k=0$) and hyperbolic ($k=-1$) structure of horizon, so the interval is given by
\begin{equation}\label{interval}
ds^2=-f(r)dt^2+\frac{dr^2}{f(r)}+r^2d\Omega_k^2,
\end{equation}
with
\begin{equation}
d\Omega_k^2=
\begin{cases}
d\theta^2+\sin^2\theta d\varphi^2,&\ \ \ \ k=1,
\\
d\theta^2+\theta^2d\varphi^2,&\ \ \ \ k=0,
\\
d\theta^2+\sinh^2\theta d\varphi^2,&\ \ \ \ k=-1,
\end{cases}
\end{equation}
where $f(r)$ is the metric function. We use an ansatz for the electromagnetic potential in the form $A_\mu=(A_t(r),0,0,A_\varphi(\theta))$ with
\begin{equation}\label{A_phi}
A_\varphi=
\begin{cases}
-q_m\cos\theta,&\ \ \ \ k=1,
\\
(1/2)q_m\theta^2,&\ \ \ \ k=0,
\\
q_m\cosh\theta,&\ \ \ \ k=-1,
\end{cases}
\end{equation}
where $q_m\geqslant0$ is the constant, which is associated with the magnetic charge. We note, that such form of electromagnetic potential gives rise to the same results for both electromagnetic field invariants $F$, $G$ and the component $A_t(r)$ of the field potential no matter what type of the geometry is (for any value of $k$), and also in this case $F$, $G$ and $A_t(r)$ do not depend on $\theta$. These facts explain the choice for $A_\varphi$ in (\ref{A_phi}). Then, there are only electric $F_{rt}(r)=dA_t/dr$ and magnetic $F_{\theta\varphi}(\theta)=dA_\varphi/d\theta$ components of electromagnetic field, namely
\begin{equation}\label{magn}
F_{\theta\varphi}=
\begin{cases}
q_m\sin\theta,&\ \ \ \ k=1,
\\
q_m\theta,&\ \ \ \ k=0,
\\
q_m\sinh\theta,&\ \ \ \ k=-1.
\end{cases}
\end{equation}
Field invariants for all $k$ are as follows
\begin{equation}
F=\frac{q_m^2-r^4F_{rt}^2}{\beta^2r^4},\qquad G=-\frac{q_mF_{rt}}{\beta^2r^2},
\end{equation}
where in the absence of the magnetic charge $q_m$ one obtains $G=0$. The electric field obtained from Eq.~(\ref{em}) for all $k$ reads
\begin{equation}\label{el}
F_{rt}=\frac{\beta q_e}{\sqrt{q_e^2+q_m^2+\beta^2r^4}},
\end{equation}
where $q_e\geqslant0$ is the electric charge of the black hole. As it can be seen the electric and magnetic charges enter into (\ref{el}) non-equivalently.

Eq.~(\ref{grav}) and Eq.~(\ref{el}) give the metric function $f(r)$
\begin{equation}\label{f}
f=k-\frac{2M}{r}-\frac{\Lambda r^2}{3}+\frac{2}{r}\int\limits_{+\infty}^r\left(\beta^2x^2-\beta\sqrt{q_e^2+q_m^2+\beta^2x^4}\right)dx,
\end{equation}
where $M$ is the ADM mass of the black hole
\begin{equation}\label{M}
M=\frac{kr_+}{2}-\frac{\Lambda r_+^3}{6}+\int\limits_{+\infty}^{r_+}\left(\beta^2x^2-\beta\sqrt{q_e^2+q_m^2+\beta^2x^4}\right)dx,
\end{equation}
where $r_+$ is the horizon of black hole, namely the largest root of $f(r_+)=0$. The integral in (\ref{f}) is the elliptic integral of the first kind and which can be expressed in terms of special functions. The black hole solution (\ref{el}), (\ref{f}) in the case of spherical horizon ($k=1$) and absence of the magnetic charge ($q_m=0$) was considered in detail in \cite{gunas12}. The limit $\beta\to+\infty$, $q_m=0$, $k=1$ recovers the RN-AdS solution \cite{kub12}
\begin{equation}
f^{(RN)}=1-\frac{2M}{r}-\frac{\Lambda r^2}{3}+\frac{q_e^2}{r^2},\qquad F_{rt}^{(RN)}=\frac{q_e}{r^2}.
\end{equation}
We note, that the metric function (\ref{f}) does not include electric $q_e$ and magnetic $q_m$ charges separately, whereas the function $f(r)$ contains quantity $\sqrt{q_e^2+q_m^2}$ \cite{gib95}, this fact means, that electric and magnetic charges enter into the metric function on equal footing. As it will be shown further, such situation takes place for the considered below thermodynamic behavior of the black hole. This fact allows us to set $q_m=0$ by plotting the graphs in these cases without loss of generality. Behavior of the function $f(r)$ and the RN-AdS ones is shown on the Fig.~\ref{solution}~(a) for various $\beta$. The electric field is shown on the Fig.~\ref{solution}~(b) for different values of the magnetic charge. It is easy to see from (\ref{el}) or Fig.~\ref{solution}~(b), that for larger values of the magnetic charge $q_m$ (when the electric charge is fixed) the electric field grows slowly near the origin ($r=0$).
\begin{figure}[ht]
\centering
\subfloat[$f(r)$ for various $\beta$]{\includegraphics[width=0.25\textwidth]{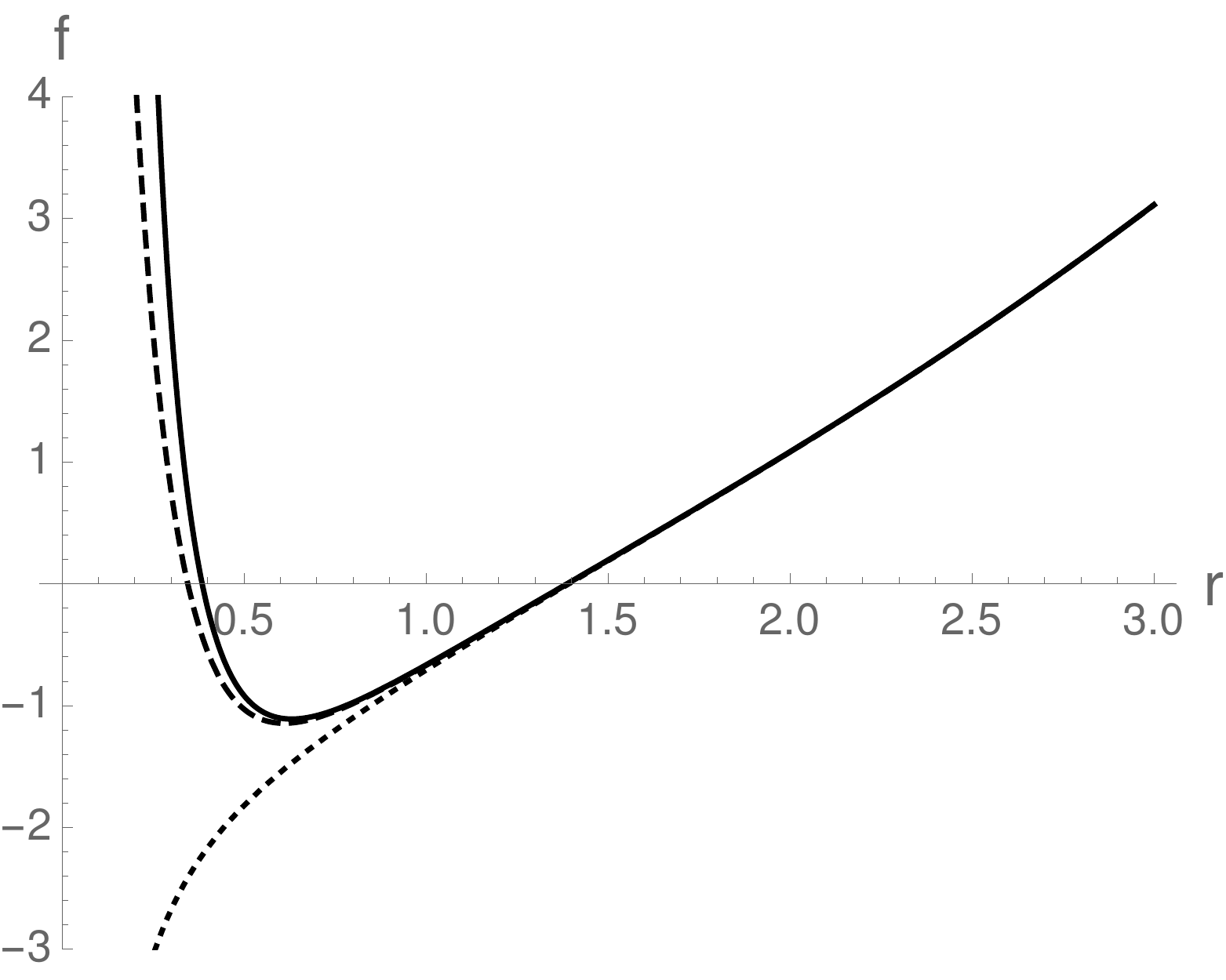}}
\qquad
\subfloat[$F_{rt}(r)$ for various $q_m$]{\includegraphics[width=0.25\textwidth]{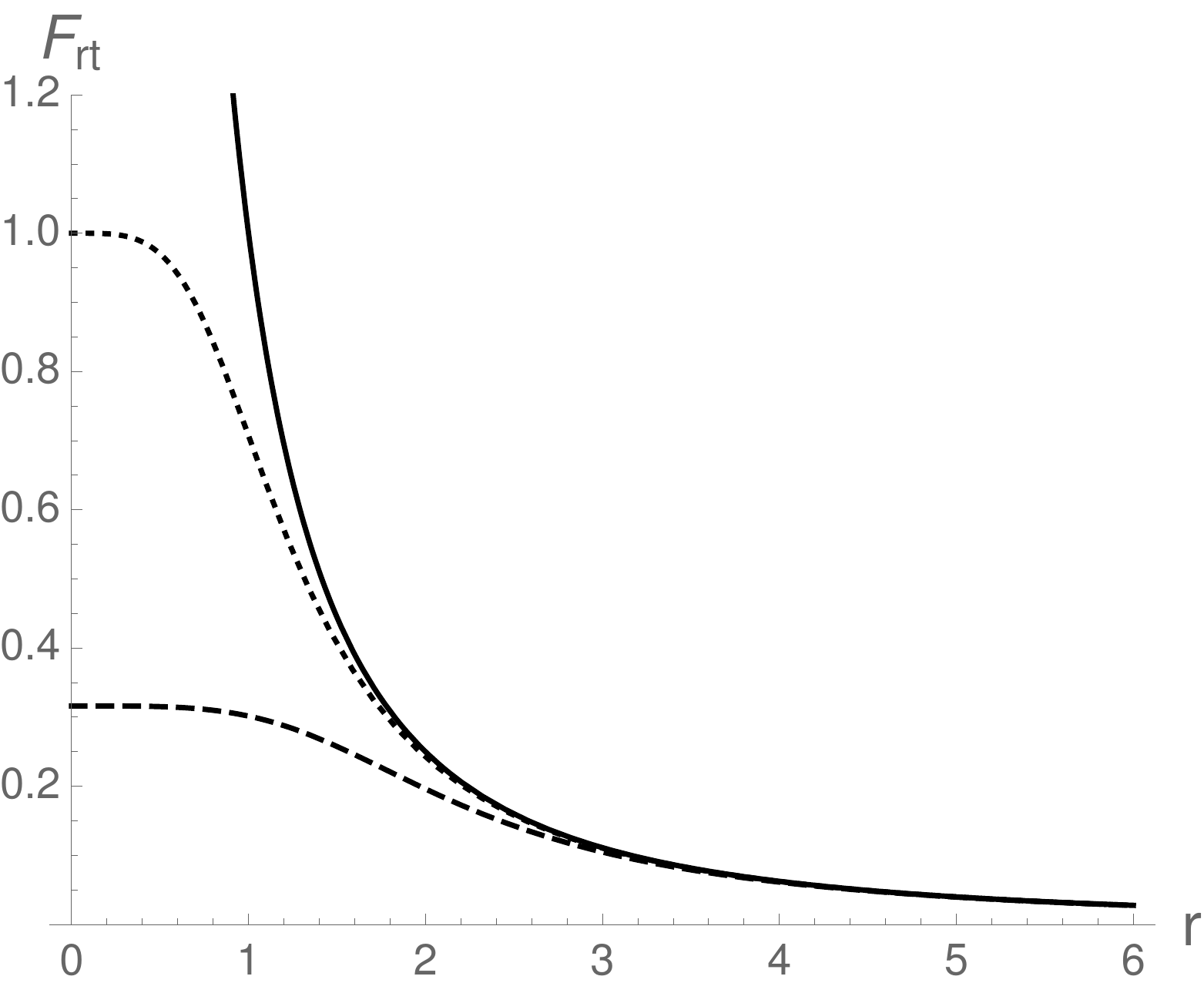}}
\caption{(a) $k=1$, $M=1.5$, $\Lambda=-1$, $q_m=0$ for $\beta=1$ (dotted), $\beta=5$ (dashed) and the RN-AdS case (solid); (b) $\beta=1$, for $q_m=0$ (dotted), $q_m=3$ (dashed) and the RN-AdS case (solid). For all graphs $q_e=1$.}
\label{solution}
\end{figure}
\section{Thermodynamics}\label{thermodynamics}
\subsection{First law and phase transition}
For chosen metric (\ref{interval}) the temperature and electric potential measured at infinity are given by $T=f'(r)/(4\pi)$ (by $r=r_+$), $U_e=-A_t(r_+)=\int_{r_+}^{+\infty}F_{rt}(r)dr$, respectively, which give
\begin{equation}\label{T}
T=\frac{1}{4\pi r_+}\left(k-\Lambda r_+^2+2\beta^2r_+^2-2\beta\sqrt{q_e^2+q_m^2+\beta^2r_+^4}\right),
\end{equation}
\begin{equation}\label{Ue}
U_e=\int\limits_{r_+}^{+\infty}\frac{\beta q_e}{\sqrt{q_e^2+q_m^2+\beta^2x^4}}dx,
\end{equation}
and for $k=1$, $q_m=0$ these quantities were considered in \cite{gunas12}. Expressions (\ref{T}) and (\ref{Ue}) reduce to the RN-AdS ones if $\beta\to+\infty$, $q_m=0$, $k=1$ \cite{kub12}
\begin{equation}
T^{(RN)}=\frac{1}{4\pi r_+}\left(1-\Lambda r_+^2-\frac{q_e^2}{r_+^2}\right),\qquad U_{e}^{(RN)}=\frac{q_e}{r_+}.
\end{equation}

A phase transition of a charged Born-Infeld-AdS black hole is admitted only for the spherical black hole horizon \cite{kub12}. Presence of the magnetic charge does not change this fact, it is easily seen from the expression for temperature (\ref{T}) where electric and magnetic charges are equivalent. Therefore, below we will consider only spherical case ($k=1$). The first law of black hole thermodynamics in the extended phase space, where the negative cosmological constant is associated with the thermodynamic pressure $P=-\Lambda/(8\pi)$ \cite{kub17} reads
\begin{equation}\label{dM}
dM=TdS+VdP+U_edq_e+U_mdq_m+Bd\beta,
\end{equation}
where $M$ is the mass function of the black hole identified with enthalpy. Using the area law one has entropy $S=A/4$, where sphere area $A$ is equal to $A=4\pi r_+^2$ and entropy is $S=\pi r_+^2$, whereas the thermodynamic volume $V$ obtained from (\ref{dM}) coincides with geometric volume of a sphere with radius $r_+$, namely $V=(4\pi/3)r_+^3$. The quantitiy $U_m$ is the magnetic potential, whereas $B$ is so-called Born-Infeld vacuum polarization \cite{wei10,gunas12}. These two terms $U_mdq_m$ and $Bd\beta$ must be included in the first law (\ref{dM}) in order to satisfy the Smarr relation in the form
\begin{equation}
M=2(TS-VP)+U_eq_e+U_mq_m-B\beta.
\end{equation}
The magnetic potential obtained from (\ref{dM}) is
\begin{equation}\label{Um}
U_m=\int\limits_{r_+}^{+\infty}\frac{\beta q_m}{\sqrt{q_e^2+q_m^2+\beta^2x^4}}dx,
\end{equation}
which coincides with the electric ones (\ref{Ue}) by replacing $q_e\to q_m$, $q_m\to q_e$. It is also in agreement with \cite{lu13}, where a magnetic potential was calculated through using the Hodge dualisation, which exchanges the roles of the electric and magnetic charges. The Born-Infeld vacuum polarization is
\begin{equation}
B=\frac{2}{3}\beta r_+^3-\frac{2}{3}r_+\sqrt{q_e^2+q_m^2+\beta^2r_+^4}+\frac{1}{3}\int\limits_{r_+}^{+\infty}\frac{q_e^2+q_m^2}{\sqrt{q_e^2+q_m^2+\beta^2x^4}}dx.
\end{equation}
We note, that in the limit $\beta\to+\infty$ the quantity $B$ is equal to zero.

Using (\ref{T}) one writes equation of state $P(v,T)$
\begin{equation}\label{P}
P=\frac{T}{v}-\frac{1}{2\pi v^2}-\frac{\beta^2}{4\pi}+\frac{\beta}{\pi v^2}\sqrt{q_e^2+q_m^2+\beta^2v^4/16},
\end{equation}
where $v=2r_+$ is the specific volume, which was introduced \cite{kub12} by dymensional analysis (translation from geometric units to physical ones) with subsequent comparison of Eq.~(\ref{P}) with the Van der Waals equation of state $P=T/(v-b)-a/v^2$. We note, that Eq.~(\ref{P}) together with its critical behavior for the case $q_m=0$ was studied in detail in \cite{gunas12}.

Taking into account some similarity between the Van der Waals equation of state and Eq.~(\ref{P}), one can check the latter one on the existence of a phase transition. The system for critical parameters $v=v_c$ and $T=T_c$ reads
\begin{equation}
\begin{cases}
(\partial P/\partial v)_T=0,
\\
(\partial^2P/\partial v^2)_T=0,
\end{cases}
\end{equation}
which gives for $v_c$ and $T_c$
\begin{equation}\label{vc}
\begin{cases}
8(q_e^2+q_m^2+\beta^2v_c^4/16)^{3/2}=\beta(q_e^2+q_m^2)[16(q_e^2+q_m^2)+3\beta^2v_c^4],
\\
T_c=\displaystyle\frac{1}{\pi v_c}\left(1-\frac{2\beta(q_e^2+q_m^2)}{\sqrt{q_e^2+q_m^2+\beta^2v_c^4/16}}\right).
\end{cases}
\end{equation}
The first equation of the system (\ref{vc}) is a cubic equation on the quantity $v_c^4$ and its analysis gives the condition between the parameter $\beta$ and charges $q_e$, $q_m$ for the existence of roots $v_c$, namely
\begin{equation}\label{inequality1}
\sqrt{q_e^2+q_m^2}>\frac{1}{\sqrt{8}\beta},
\end{equation}
which coincides with corresponding condition obtained in \cite{gunas12} in the absence of the magnetic charge. We note, that critical parameters $P_c$, $v_c$, $T_c$ depend besides $q_e$ and $q_m$ also on $\beta$. In the region
\begin{equation}
\frac{1}{\sqrt{8}\beta}<\sqrt{q_e^2+q_m^2}<\frac{1}{2\beta}
\end{equation}
there exists two real positive roots for the critical value of the specific volume $v_c$, whereas for $\sqrt{q_e^2+q_m^2}>1/(2\beta)$ only one (see Fig.~\ref{appendix}~(a) in the appendix). We also note, that on the graphs below we take the largest root for $v_c$. In the limit $\beta\to+\infty$, $q_m=0$, $k=1$ the equation of state for the RN-AdS black hole is recovered \cite{kub12}
\begin{equation}\label{P_RN}
P^{(RN)}=\frac{T}{v}-\frac{1}{2\pi v^2}+\frac{2q_e^2}{\pi v^4},
\end{equation}
together with its critical parameters
\begin{equation}
P_c^{(RN)}=\frac{1}{96\pi q_e^2},\qquad v_c^{(RN)}=2\sqrt{6}q_e,\qquad T_c^{(RN)}=\frac{\sqrt{6}}{18\pi q_e}.
\end{equation}
Eq.~(\ref{P_RN}) admits a phase transition for arbitrary positive values of the charge $q_e$, and this is consistent with inequality (\ref{inequality1}) as far as when $\beta\to+\infty$ and $q_m=0$ one has the condition $q_e>0$ for the existence of the phase transition. Also, Eq.~(\ref{P_RN}) expressed in terms of $p=P/P_c$, $\nu=v/v_c$, $\tau=T/T_c$ does not depend on parameter $q_e$ as for the Van der Waals gas, whereas Eq.~(\ref{P}) does not have such feature. We remark that Eq.~(\ref{P}) at large specific volume $v$ gives the ideal gas asymptotic $P=T/v$ for finite $\beta$ as for Eq.~(\ref{P_RN}). The critical ratio reads
\begin{equation}\label{ratio}
\frac{P_cv_c}{T_c}=1-\frac{2+\beta^2v_c^2-4\beta\sqrt{q_e^2+q_m^2+\beta^2v_c^4/16}}{4-8\beta(q_e^2+q_m^2)(q_e^2+q_m^2+\beta^2v_c^4/16)^{-1/2}},
\end{equation}
where the value of $v_c$ is given by Eq.~(\ref{vc}). In the RN-AdS limit this ratio recovers the universal critical ratio of the RN-AdS black hole \cite{altam14}
\begin{equation}
\frac{P_cv_c}{T_c}=\frac{3}{8},
\end{equation}
as for the Van der Waals system. We point out that an interesting mathematical feature of the ratio (\ref{ratio}) (and hence Eq.~(\ref{P})) is its dependence only on the product $\beta\sqrt{q_e^2+q_m^2}$ and thereby this ratio remains unchanged for a fixed value of $\beta\sqrt{q_e^2+q_m^2}$. This fact is demonstrated on the Fig.~\ref{appendix2} in the appendix.

\begin{figure}[ht]
\centering
\subfloat[critical isotherms for various $\beta$]{\includegraphics[width=0.25\textwidth]{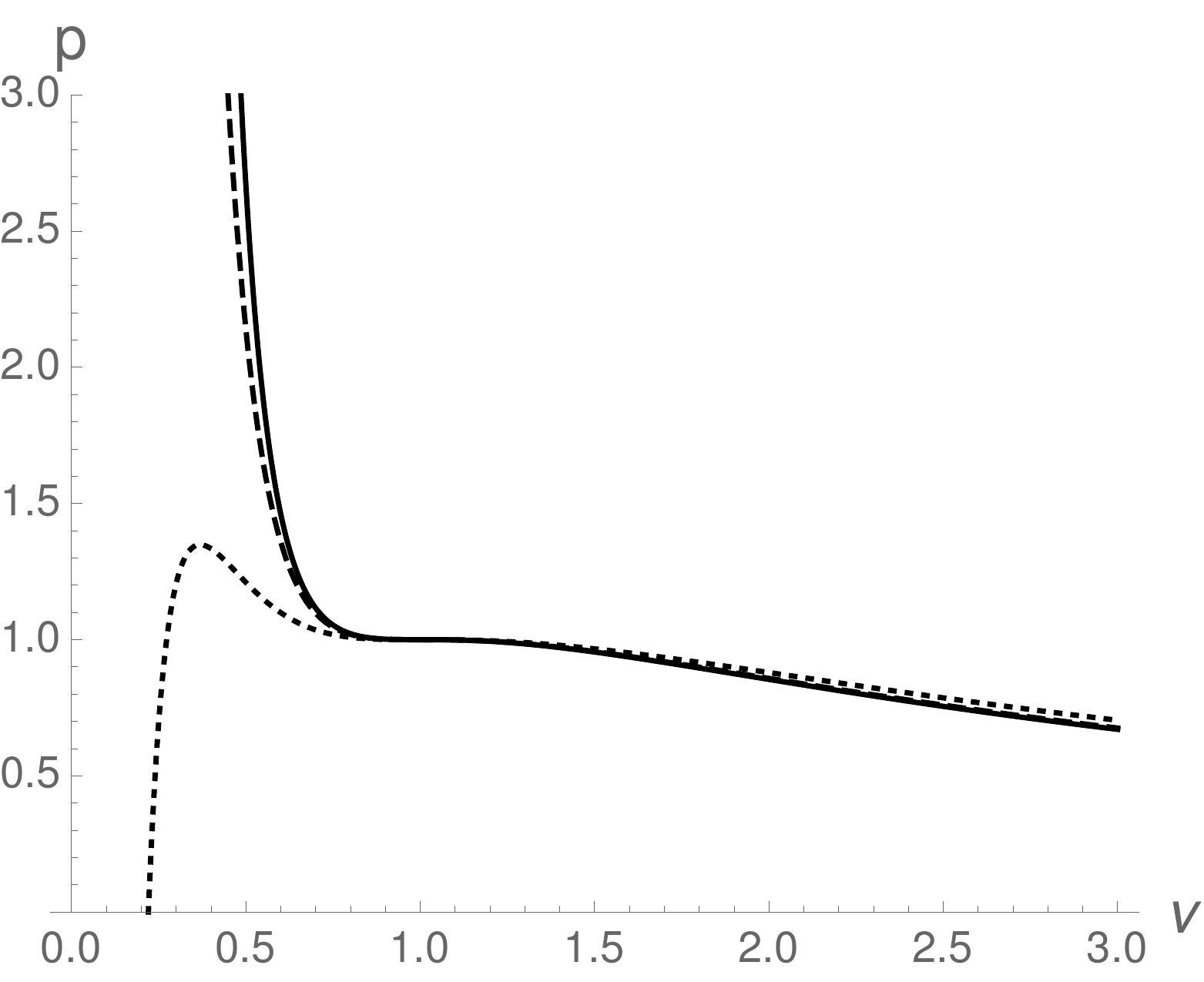}}
\qquad
\subfloat[critical isotherms for various $q_e$]{\includegraphics[width=0.25\textwidth]{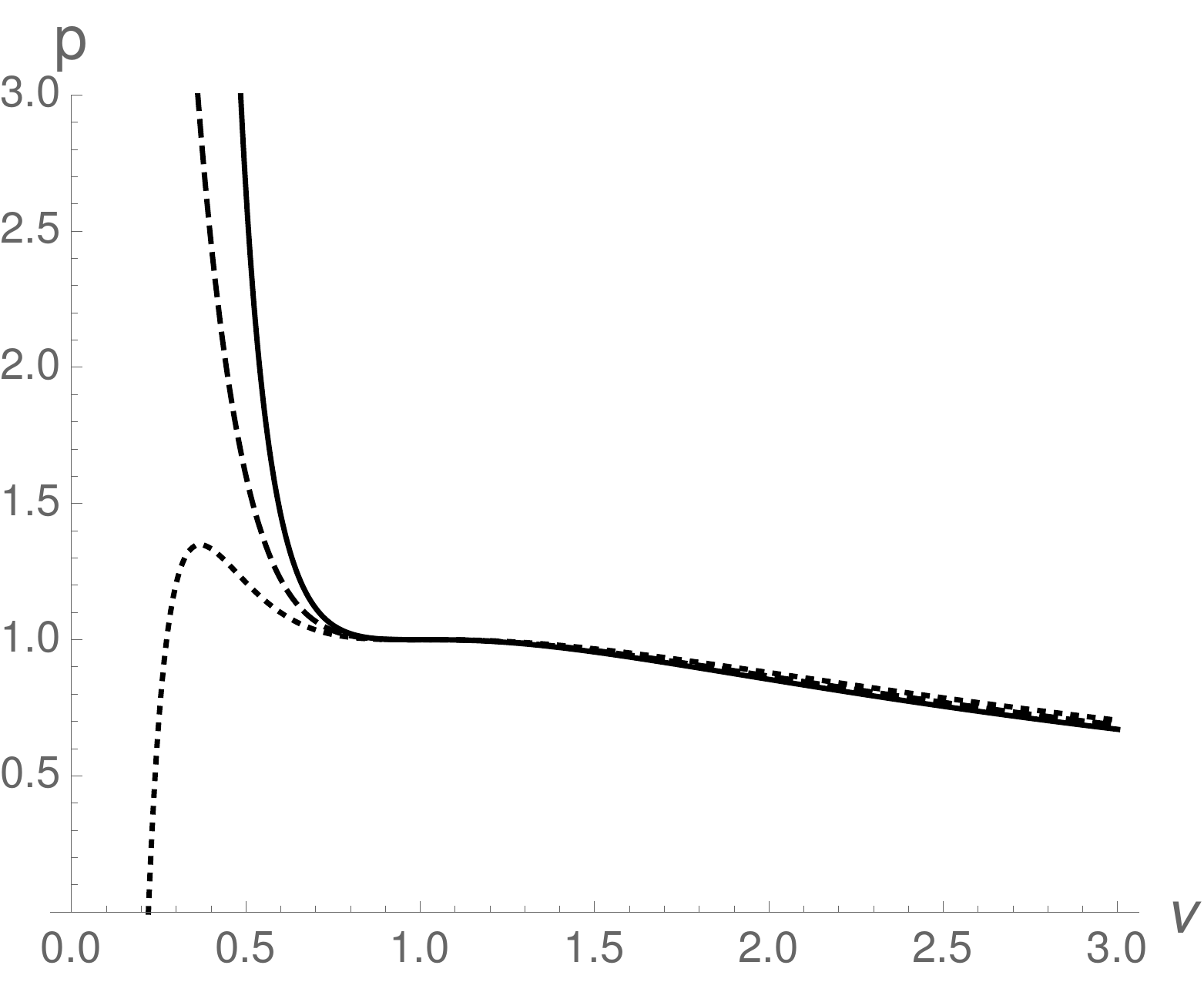}}
\caption{(a) $q_e=1$, for $\beta=0.4$ (dotted), $\beta=0.8$ (dashed) and the RN-AdS case (solid); (b) $\beta=0.5$ for $q_e=0.8$ (dotted), $q_e=1$ (dashed), $q_e=10$ (solid). For all graphs $\tau=1$, $q_m=0$.}
\label{p(nu)_beta_qe}
\end{figure}

\begin{figure}[ht]
\centering
\subfloat[isotherms for various $\tau$]{\includegraphics[width=0.25\textwidth]{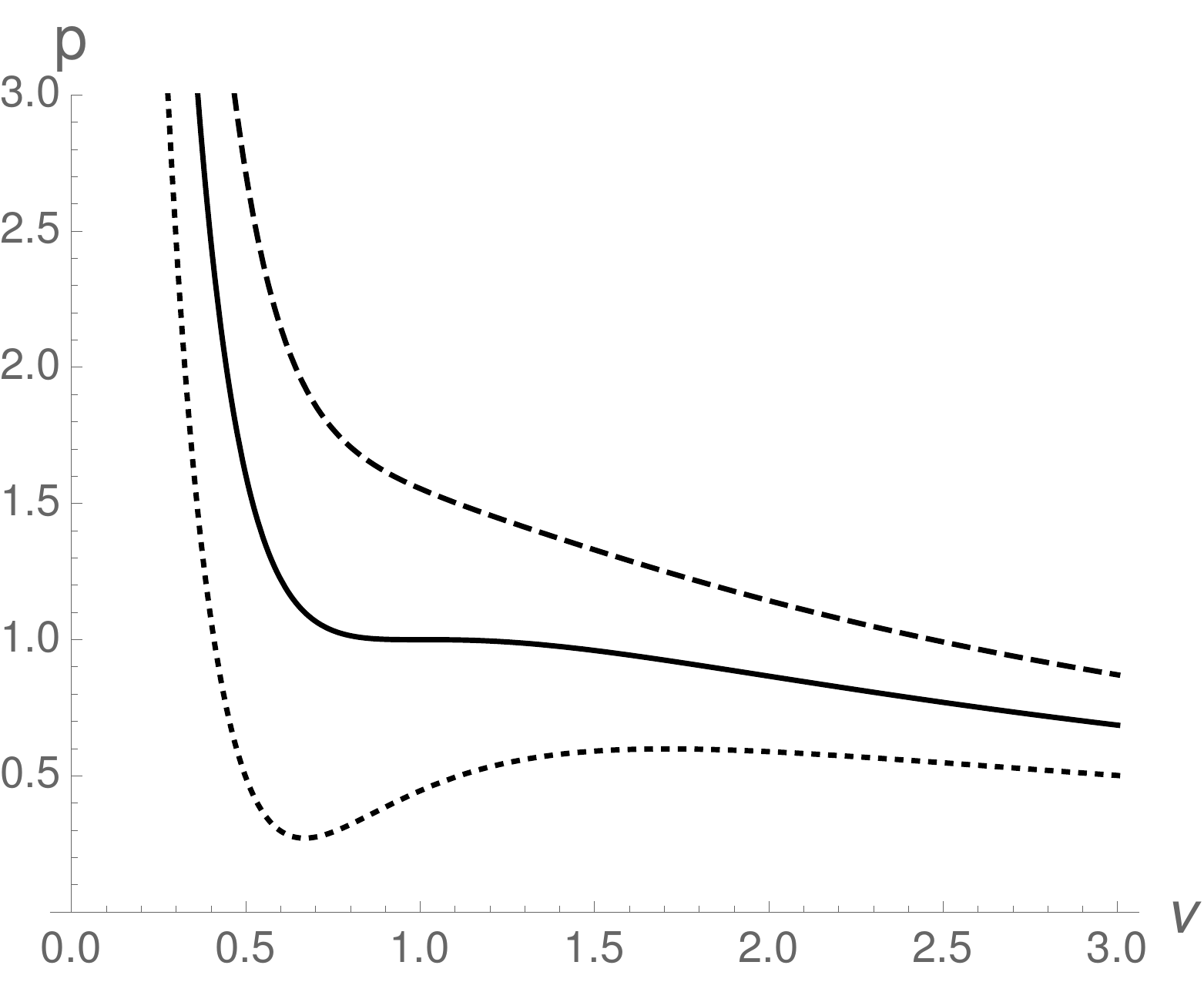}}
\qquad
\subfloat[heat capacity for various $\beta$]{\includegraphics[width=0.25\textwidth]{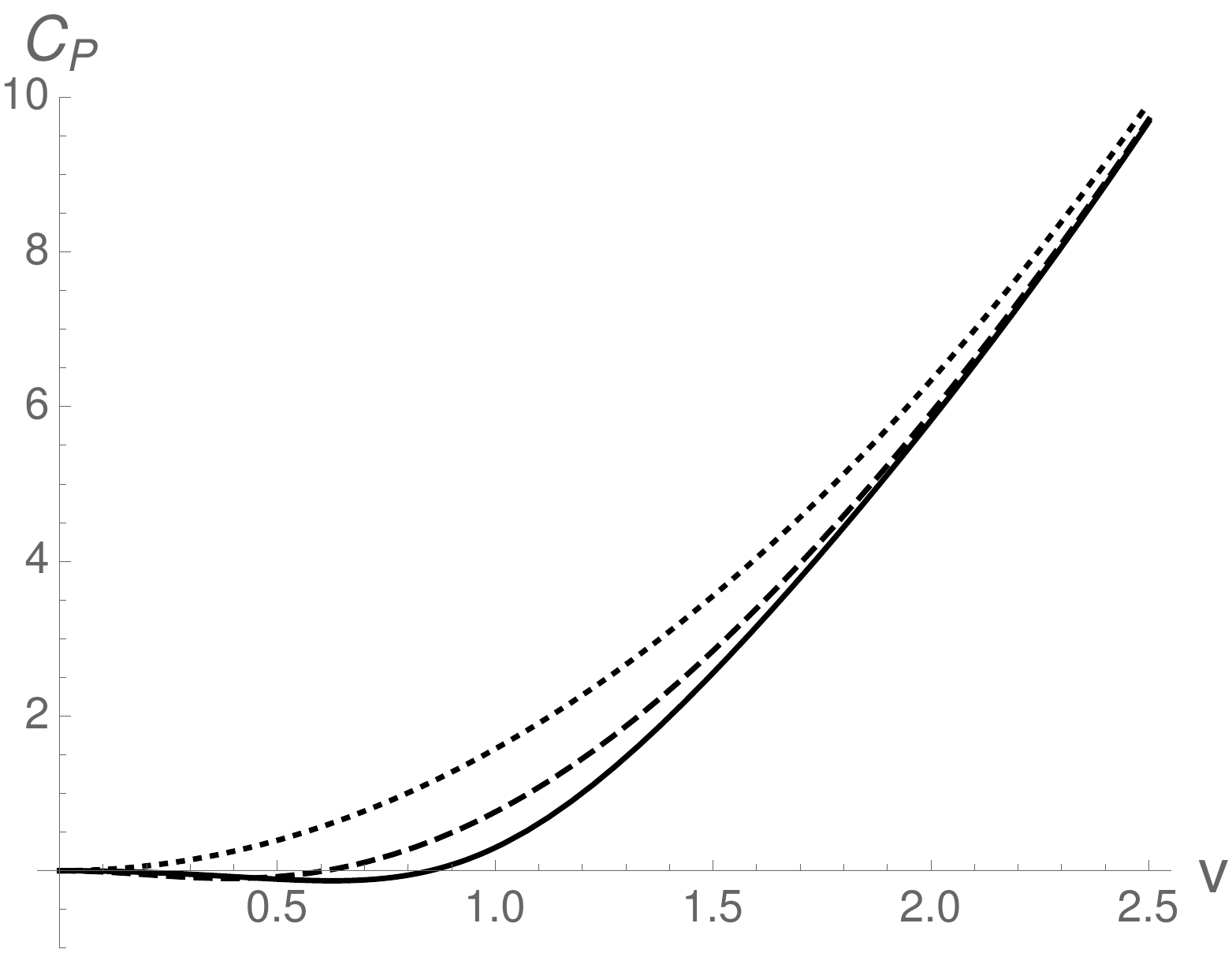}}
\caption{(a) $\beta=0.5$ for $\tau=0.8$ (dotted), $\tau=1$ (solid), $\tau=1.2$ (dashed); (b) $P=1$ for $\beta=0.5$ (dotted), $\beta=2$ (dashed) and the RN-AdS case (solid). For all graphs $q_e=1$, $q_m=0$.}
\label{p(nu)_beta_qe_CP(v)_beta}
\end{figure}

At the end of this subsection we give the heat capacity at constant pressure $C_P=T(\partial S/\partial T)_P$ for spherical horizon
\begin{equation}
C_P=\frac{\pi v^2}{2}\frac{\left(2+4\pi Pv^2+\beta^2v^2-4\beta\sqrt{q_e^2+q_m^2+\beta^2v^4/16}\right)}{\left(-2+4\pi Pv^2+\beta^2v^2+\displaystyle\frac{4\beta(q_e^2+q_m^2-\beta^2v^4/16)}{\sqrt{q_e^2+q_m^2+\beta^2v^4/16}}\right)},
\end{equation}
which coincides with \cite{bi21} when $q_m=0$. The RN-AdS limit leads to \cite{kub12}
\begin{equation}
C_P^{(RN)}=\frac{\pi v^2}{2}\frac{(v^2+2\pi Pv^4-4q_e^2)}{(-v^2+2\pi Pv^4+12q_e^2)}.
\end{equation}
Critical isotherms on the $p - \nu$ plane for various values of the parameter $\beta$ for fixed charges $q_e$, $q_m$, and for various values of the electric charge $q_e$ for fixed $\beta$ are shown on the Fig.~\ref{p(nu)_beta_qe}~(a,b), respectively. These graphs demonstrate the dependence of the equation of state (\ref{P}) in terms $p$, $\nu$, $\tau$ on parameters $\beta$ and $q_e$, respectively. Isotherms on the $p - \nu$ plane for different values of temperature $\tau$ for finite $\beta$, and the heat capacity $C_P$ for various $\beta$ are shown on the Fig.~\ref{p(nu)_beta_qe_CP(v)_beta}~(a,b), respectively. In addition, solutions of the first equation of the system (\ref{vc}) versus $\beta$ for various $q_e$ are shown on the Fig.~\ref{appendix}~(a) (in the appendix), and this graph illustrates the inequality (\ref{inequality1}), particularly the region $1/(\sqrt{8}\beta)<\sqrt{q_e^2+q_m^2}<1/(2\beta)$, where there are two real positive roots for $v_c$. The analogous graph is presented in paper \cite{gunas12}.
\subsection{The Joule-Thomson expansion}
\begin{figure}[ht]
\centering
\subfloat[inversion curves for various $\beta$]{\includegraphics[width=0.25\textwidth]{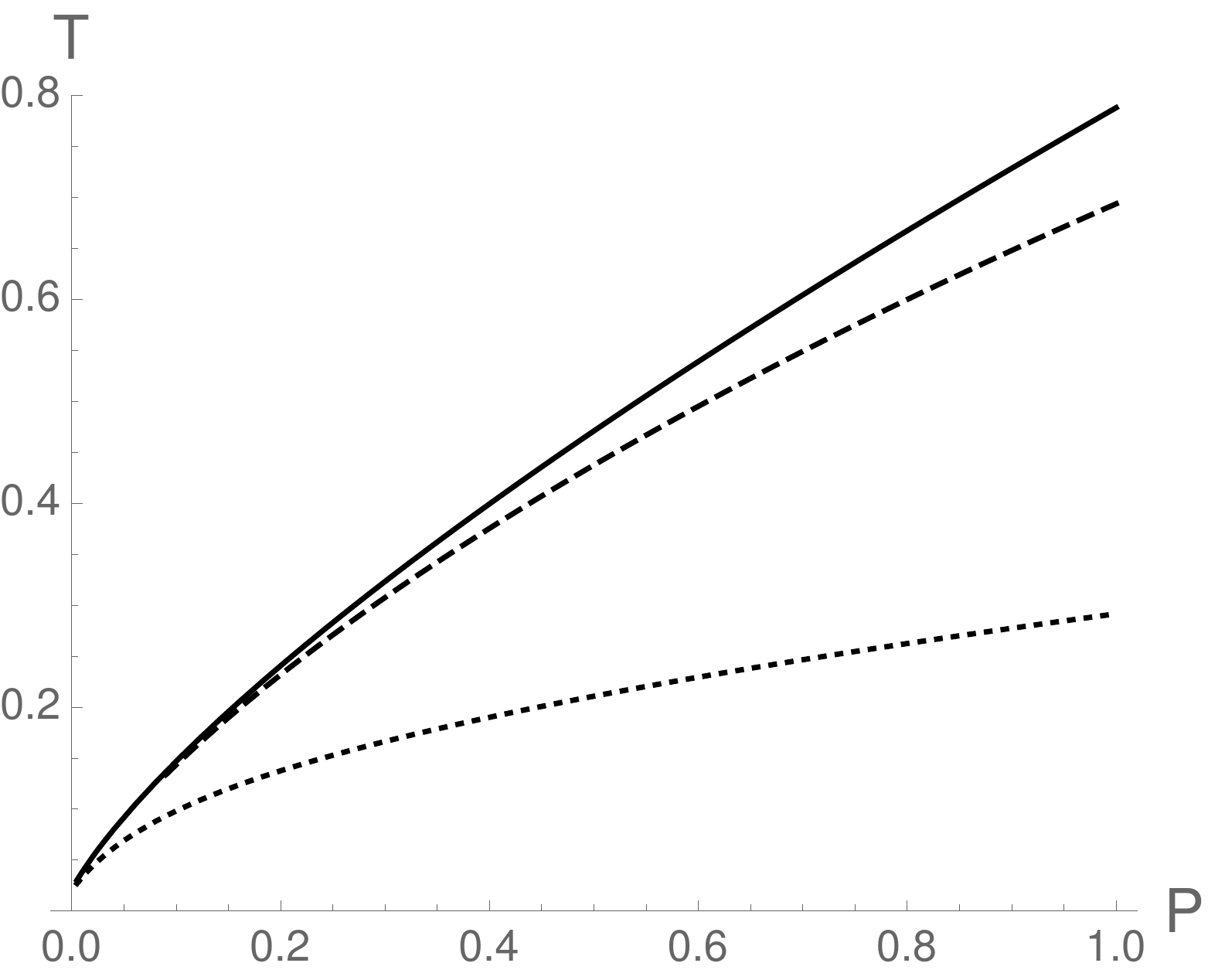}}
\qquad
\subfloat[isenthalpic curves for various $M$]{\includegraphics[width=0.25\textwidth]{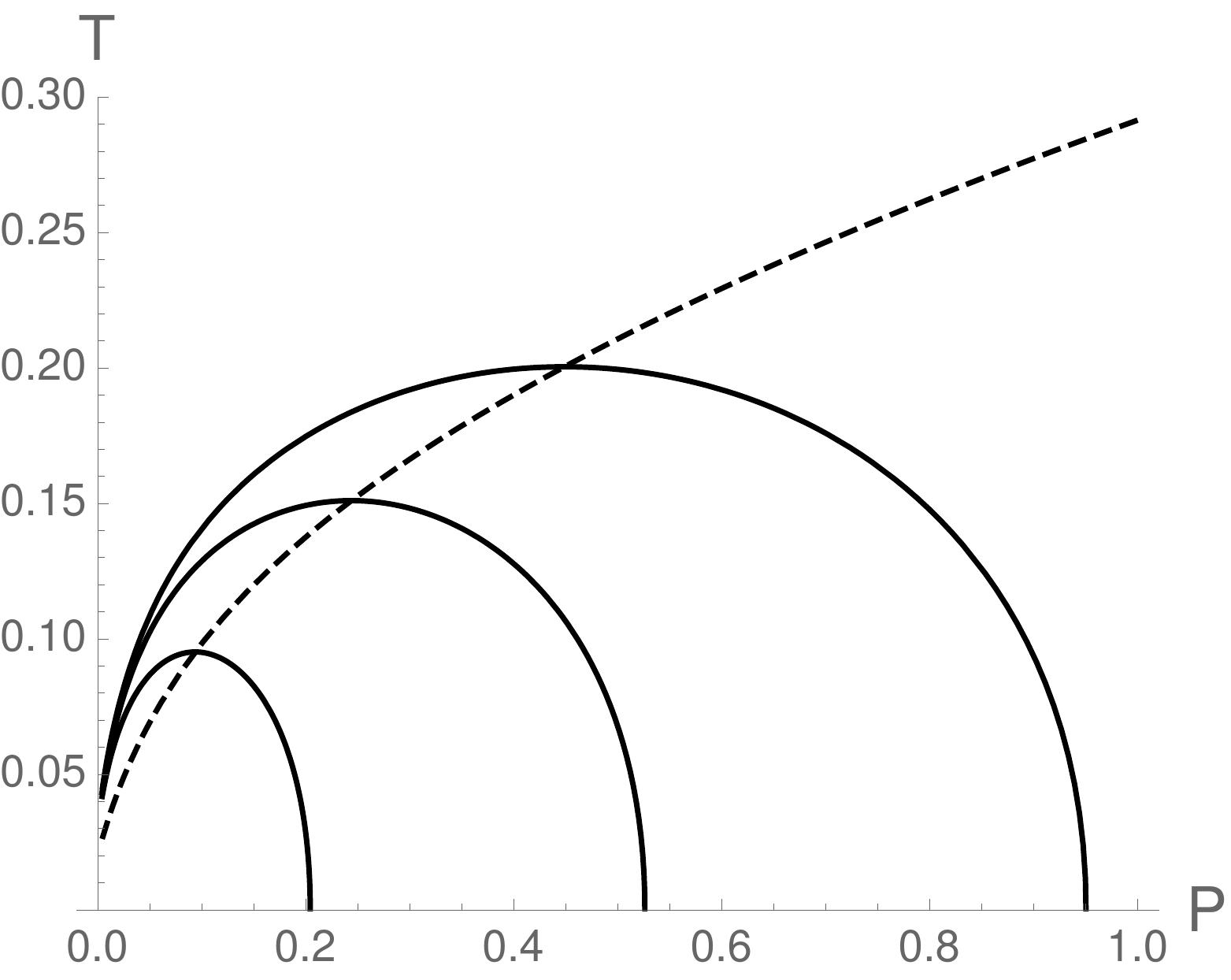}}
\caption{(a) $\beta=1$ (dotted), $\beta=4$ (dashed) and the RN-AdS case (solid); (b) $\beta=1$ for $M=1.11,1.15,1.17$ (from bottom to top). The dashed line marks the inversion curve with the same parameters. For all graphs $q_e=1$, $q_m=0$.}
\label{T(P)}
\end{figure}

Association of the negative cosmological constant with the thermodynamic pressure means, that the black hole is considered to be immersed in the environment which can be effectively described by a negative cosmological constant, thus we consider the pressure of this environment on the black hole \cite{kub17}. Such approach allows to investigate critical behavior, heat cycles, compressibility, Joule-Thomson expansion of black holes \cite{okcu17}. Let us consider the latter of these in more detail for spherical horizon. The Joule-Thomson coefficient reads
\begin{equation}\label{mu}
\mu=\left(\frac{\partial T}{\partial P}\right)_M=\frac{T(\partial V/\partial T)_P-V}{C_P}=-\frac{T(\partial P/\partial T)_V+V(\partial P/\partial V)_T}{C_P(\partial P/\partial V)_T},
\end{equation}
as far as the mass of black hole is identified with enthalpy, and here $V$ is the thermodynamic volume. At the inversion temperature $T_i$ and pressure $P_i$ the coefficient $\mu=0$ and this determines the cooling ($\mu>0$) and heating ($\mu<0$) regions on the $T - P$ plane. During the expansion by constant enthalpy (black hole mass) the thermodynamic system cools down for $\mu>0$ and warms up for $\mu<0$, this is effect of the Joule-Thomson expansion. Taking into account the relation between the thermodynamic volume and specific volume, and equating the right hand side of Eq.~(\ref{mu}) to zero one obtains the system of relations for inversion values $T_i$ and $P_i$
\begin{equation}\label{JT}
\begin{cases}
T_i=\displaystyle-\frac{1}{2\pi v}+\frac{\beta(q_e^2+q_m^2)}{\pi v\sqrt{q_e^2+q_m^2+\beta^2v^4/16}},
\\
P_i=\displaystyle\frac{T_i}{v}-\frac{1}{2\pi v^2}-\frac{\beta^2}{4\pi}+\frac{\beta}{\pi v^2}\sqrt{q_e^2+q_m^2+\beta^2v^4/16},
\end{cases}
\end{equation}
and the condition on $v$ (thereby on the black hole horizon, since $v=2r_+$) for the existence of the solution of this system is
\begin{equation}\label{quartic}
1+\pi P_iv^2+\frac{\beta^2v^2}{4}-\frac{\beta[2(q_e^2+q_m^2)+\beta^2v^4/16]}{\sqrt{q_e^2+q_m^2+\beta^2v^4/16}}=0.
\end{equation}
Eq.~(\ref{quartic}) is a quartic equation on the quantity $v^2$, and for $v$ one has one real positive root of Eq.~(\ref{quartic}) (see Fig.~\ref{appendix}~(b,c) in the appendix). Analysis of Eq~(\ref{quartic}) gives a condition of the existence such root as
\begin{equation}\label{inequality2}
\sqrt{q_e^2+q_m^2}>\frac{1}{2\beta},
\end{equation}
and this condition is the same as for the existence of one real positive root $v_c$ for phase transition, we also note that this condition does not depend on $P$. For the RN-AdS black hole for $\beta\to+\infty$, $q_m=0$ one has a condition $q_e>0$ and the inversion curve $T_i(P_i)$ is given by \cite{okcu17}
\begin{equation}
T_i=\frac{\sqrt{P_i}}{\sqrt{2\pi}}\frac{\left(1+16\pi q_e^2P_i-\sqrt{1+24\pi q_e^2P_i}\right)}{\left(-1+\sqrt{1+24\pi q_e^2P_i}\right)^{3/2}}.
\end{equation}
As it was noted earlier, the black hole mass is identified with enthalpy, thereby using the black hole mass $M(r_+,\Lambda)$ of Eq.~(\ref{M}) and the temperature $T(r_+,\Lambda)$ of Eq.~(\ref{T}) with $P=-\Lambda/(8\pi)$ and setting $k=1$ one obtains the isenthalpic curves $T(P,M)$ for fixed black hole mass. Inversion curves on the $T - P$ plane for various $\beta$, and isenthalpic curves for different values of the black hole mass $M$ for finite $\beta$ are shown on the Fig.~\ref{T(P)}~(a,b), respectively. On the Fig.~\ref{T(P)}~(b) above the inversion curve one has the cooling region ($\mu>0$, system cools down by isenthalpic expansion), and below the inversion curve there is the heating region ($\mu<0$, system warms up by isenthalpic expansion). Also,  solutions of Eq.~(\ref{quartic}) versus $\beta$ for various $q_e$ and $P$ are shown on the Fig.~\ref{appendix}~(b,c) (in the appendix), this graphs illustrates the inequality (\ref{inequality2}). We should notice, that similar results were obtained by authors of a recent paper \cite{bi21} for electrically charged Born-Infeld-AdS black holes.
\section{Conclusions}\label{conclusions}
In this paper we have considered the electrically and magnetically charged static black hole solution with the Born-Infeld Lagrangian in the four-dimensional anti-de Sitter spacetime with spherical, planar and hyperbolic horizon structures. An ansatz for the electromagnetic field potential was chosen in the form which gives rise to the same form of electric field and field invariants for all types of considered horizon geometries. The electric field $F_{rt}$ grows slowly near the origin for larger values of the magnetic charge for fixed the electric one and is finite at $r=0$.

Temperature, electric and magnetic potentials were calculated and they satisfy the first law of black hole thermodynamics in extended phase space where the negative cosmological constant plays a role of the thermodynamic pressure. The Smarr relation is valid when one considers the additional term $Bd\beta$ in the first law, where $B$ stands for the Born-Infeld vacuum polarization. The phase transition is admitted only for the spherical topology, and corresponding conditions for its existence were considered. The critical ratio depends only on the product $\beta\sqrt{q_e^2+q_m^2}$ and it shows its universality. Also the heat capacity by constant pressure was obtained and investigated, namely for a chosen range of parameter $\beta$ heat capacity is a monotonous function for intermediate values of volume.

In the absence of the magnetic charge ($q_m=0$), solutions and their thermodynamic properties considered in this article reduce to the results previously obtained in the literature \cite{gunas12,okcu17} for a electrically charged Born-Infeld-AdS black hole. We point out that electric and magnetic charges enter into the metric function symmetrically and from the qualitative point of view when we set $q_e=0$ or $q_m=0$ the behavior of the metric function as well as all the consequent notions would be identical. It should be also noted that in the limit $q_m=0$ the magnetic potential (and magnetic field) disappears. The limit $\beta\to+\infty$ leads to the RN-AdS black hole what is pretty expected.

That fact that the black hole mass is identified with enthalpy in extended phase space allows to consider the Joule-Thomson expansion during which the black hole mass remains unchanged. We obtained the condition $\sqrt{q_e^2+q_m^2}>1/(2\beta)$ for the existence of the Joule-Thomson expansion and in this case we plotted inversion and isenthalpic curves where the cooling and heating regions are demonstrated. With increasing of the parameter $\beta$ corresponding inversion curves assymptotically tend to the RN-AdS one from below. In the limit $\beta\to+\infty$ together with $q_m=0$ we recover already well-known in literature the Joule-Thomson expansion for the RN-AdS black hole. For a purely electric field our results stand together with ones obtained in a recent paper \cite{bi21}.

Thereby, original feature of this paper is the study of the critical behavior, including a critical ratio and the heat capacity. We also point out that the Joule-Thomson expansion for black hole with electric and magnetic charges in original Einstein-Born-Infeld theory was studied for the first time in this work. The general features of the inversion curves are very similar to their counterparts in the RN-AdS case, but in comparison with the latter one in the Born-Infeld case the inversion temperatures become lower, therefore it can be stated that presence of the Born-Infeld field gives rise to decrease of the inversion temperatures. 
\section*{Appendix}
\begin{figure}[ht]
\centering
\subfloat[solutions of the first equation of the system (\ref{vc}) for various $q_e$]{\includegraphics[width=0.25\textwidth]{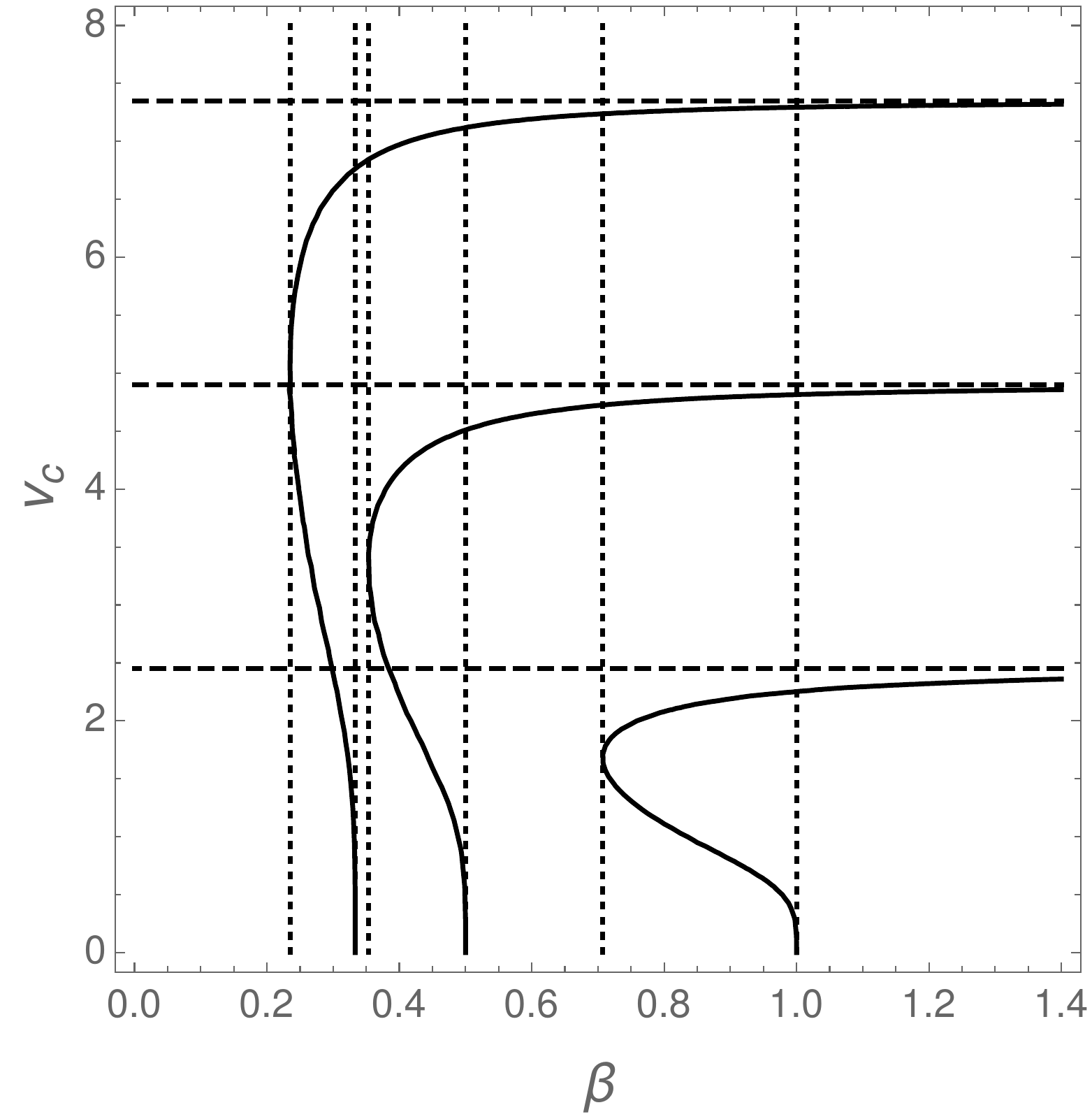}}
\qquad
\subfloat[solutions of Eq.~(\ref{quartic}) for various $q_e$ with $P=1$]{\includegraphics[width=0.25\textwidth]{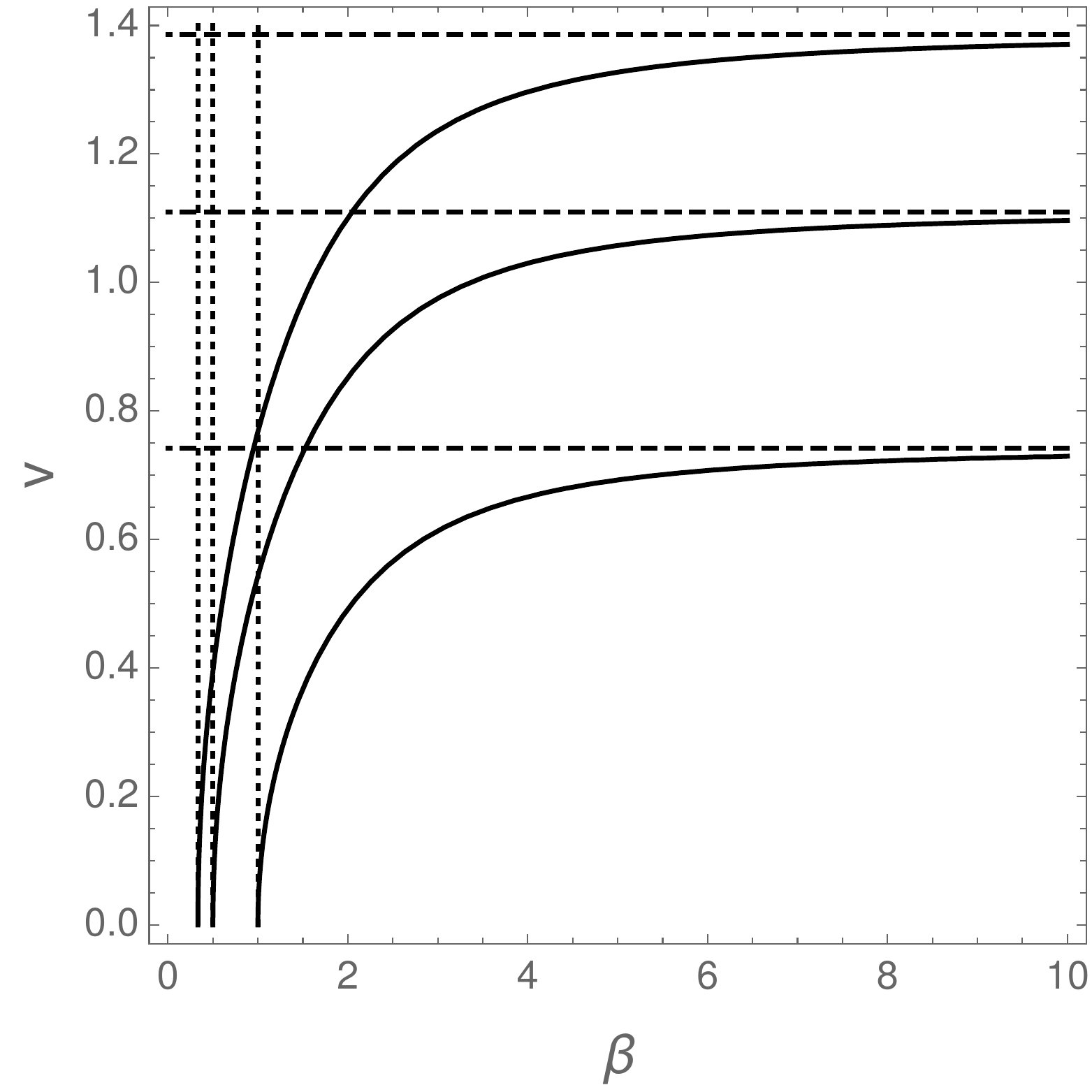}}
\qquad
\subfloat[solutions of Eq.~(\ref{quartic}) for various $q_e$ with $P=5$]{\includegraphics[width=0.25\textwidth]{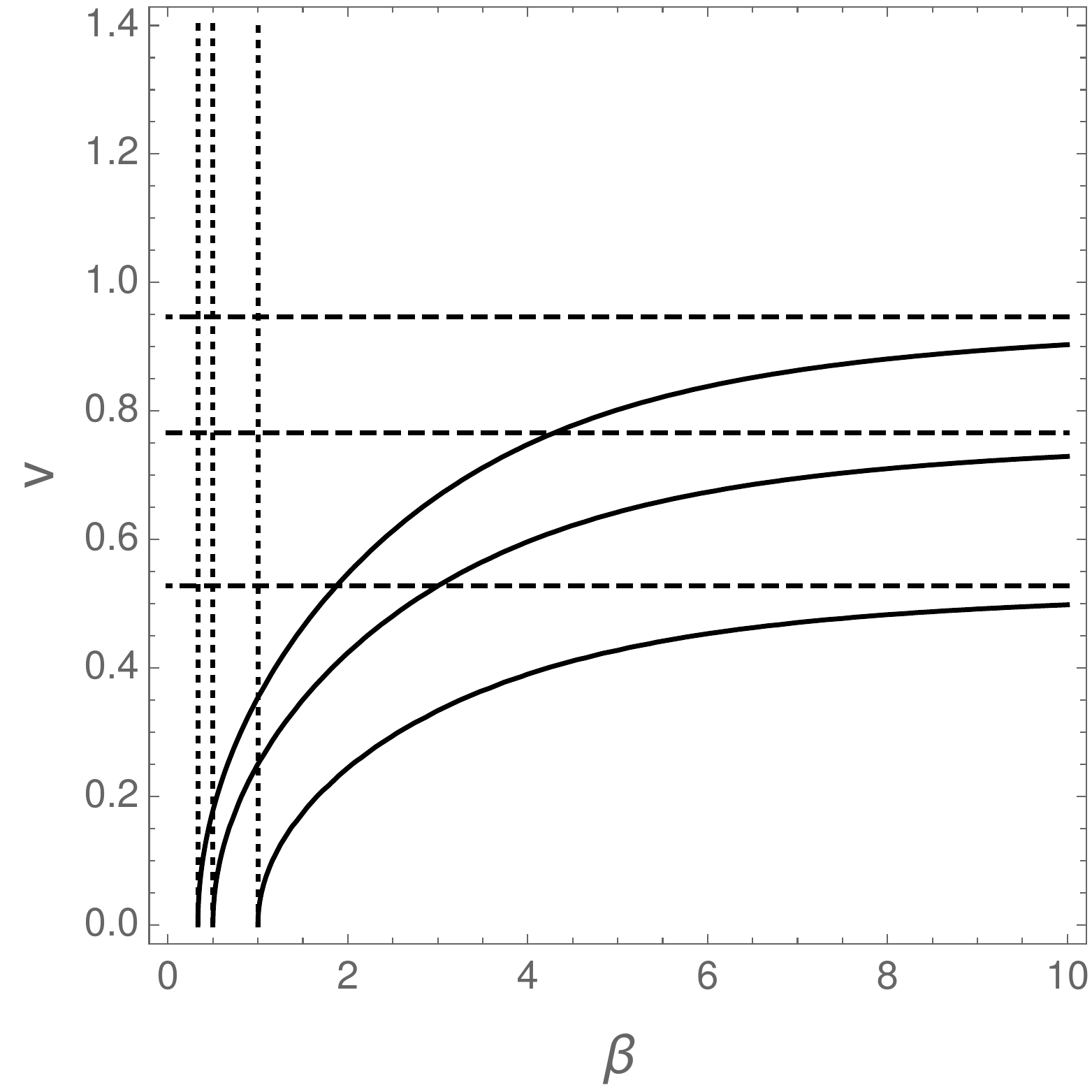}}
\caption{For all graphs $q_m=0$, $q_e=0.5,1,1.5$ (solid lines from bottom to top). Dashed lines show the RN-AdS case. Dotted lines show the equality case of (\ref{inequality1}) and the region of two real positive roots for (a) and the equality case of (\ref{inequality2}) for (b), (c).}
\label{appendix}
\end{figure}

\begin{figure}[ht]
\centering
\subfloat[critical ratio depending on $\beta$ for various $q$]{\includegraphics[width=0.25\textwidth]{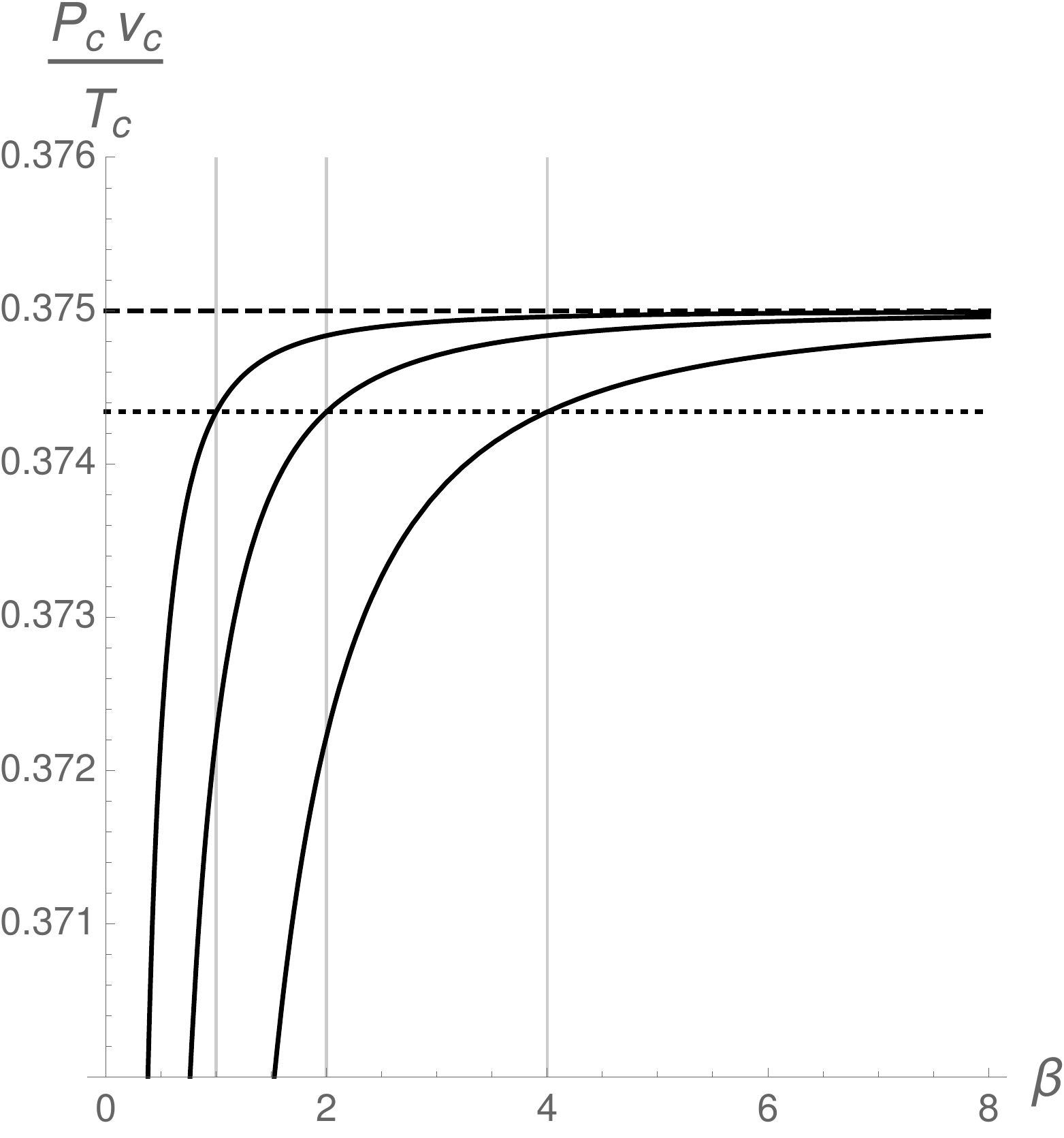}}
\qquad
\subfloat[critical ratio depending on $q$ for various $\beta$]{\includegraphics[width=0.25\textwidth]{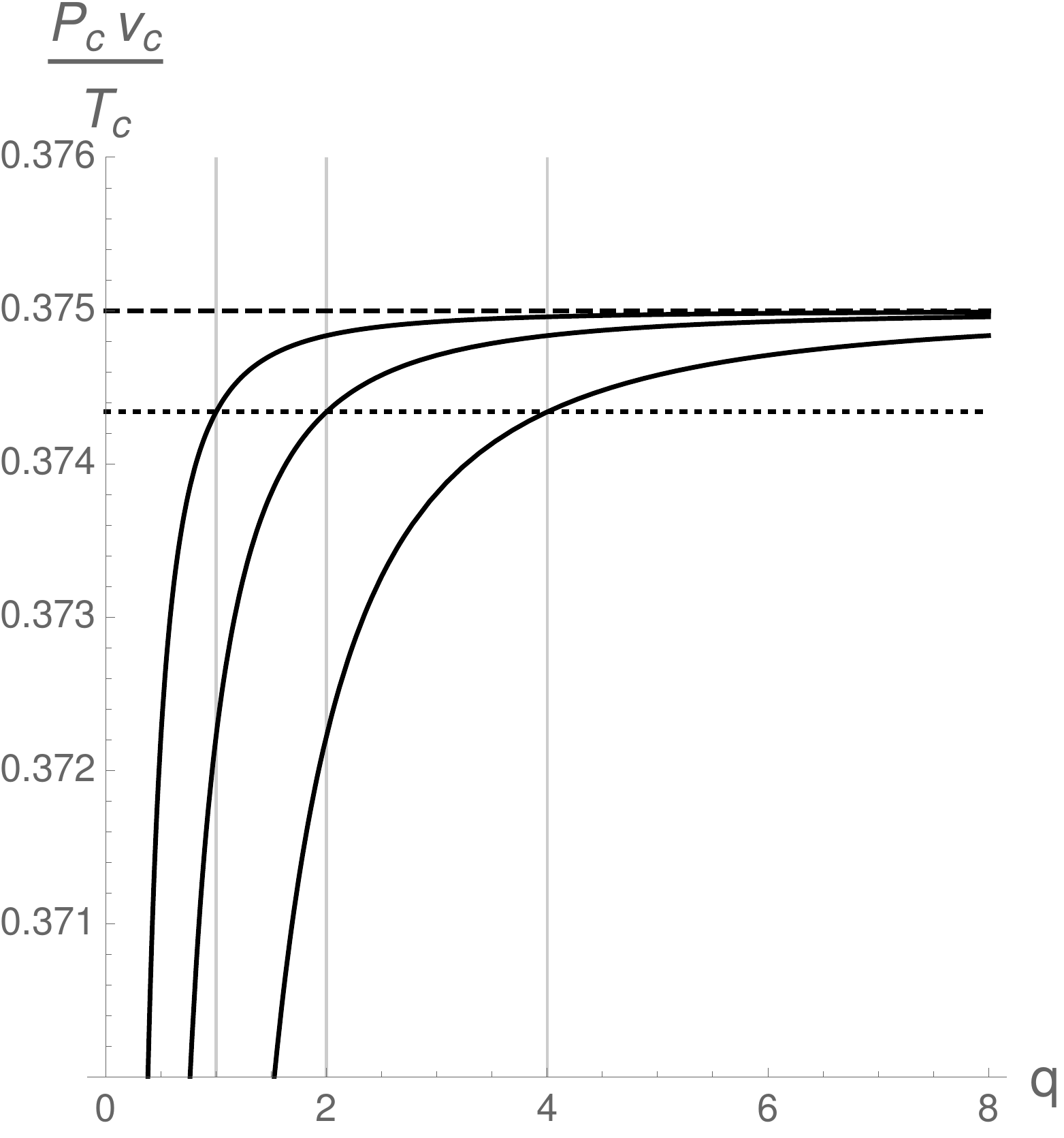}}
\caption{Here $q$ stands for $\sqrt{q_e^2+q_m^2}$. These two identical graphs demonstrate, that the critical ratio (\ref{ratio}) depends only on the combination $\beta q$. On the graph (a) solid curves correspond to $q_1=0.5$, $q_2=1$, $q_3=2$ from bottom to top. The horizontal dashed line denotes the RN-AdS limit, namely $3/8$. Vertical lines correspond to $\beta_1=4$, $\beta_2=2$, $\beta_3=1$ from right to left. The horizontal dotted line crosses the solid curves in points where $\beta_1q_1=\beta_2q_2=\beta_3q_3=2$. Parameters on the graph (b) correspond to the ones on the graph (a) by replacing $\beta\to q$, $q\to\beta$.}
\label{appendix2}
\end{figure}

Graphs on the Fig.~\ref{appendix} illustrate solutions of the first equation of the system (\ref{vc}) and also Eq.~(\ref{quartic}) depending on parameter $\beta$ for various other parameters. Graphs on the Fig.~\ref{appendix2} demonstrate the dependence of the critical ratio (\ref{ratio}) on parameters $\beta$ and charges $q_e$, $q_m$.


\begin{thebibliography}{00}
\bibitem{kas09}
D. Kastor, S. Ray, J. Traschen, Class. Quantum Grav. 26 (2009) 195011.
\bibitem{dol11}
B. P. Dolan, Class. Quantum Grav. 28 (2011) 125020.
\bibitem{dol11_2}
B. P. Dolan, Class. Quantum Grav. 28 (2011) 235017.
\bibitem{sek06}
Y. Sekiwa, Phys. Rev. D 73 (2006) 084009.
\bibitem{cvet11}
M. Cveti{\v c}, G. W. Gibbons, D. Kubiz{\v n}{\'a}k, C. N. Pope, Rhys. Rev. D 84 (2011) 024037.
\bibitem{kub17}
D. Kubiz{\v n}{\'a}k, R. B. Mann, M. Teo, Class. Quantum Grav. 34 (2017) 063001.
\bibitem{kub12}
D. Kubiz{\v n}{\'a}k, R. B. Mann, J. High Energ. Phys. 07 (2012) 033.
\bibitem{altam14}
N. Altamirano, D. Kubiz{\v n}{\'a}k, R. B. Mann, Z. Sherkatghanad, Galaxies 2 (2014) 89.
\bibitem{okcu17}
{\"O}. {\"O}kc{\"u}, E. Ayd{\i}ner, Eur. Phys. J. C 77 (2017) 24.
\bibitem{mo18}
Jie-Xiong Mo, Gu-Qiang Li, Shan-Quan Lan, Xiao-Bao Xu, Phys. Rev. D 98 (2018) 124032.
\bibitem{cist19}
A. Cisterna, Shi-Qian Hu, Xiao-Mei Kuang, Phys. Lett. B 797 (2019) 134883.
\bibitem{bi21}
S. Bi, M. Du, J. Tao, F. Yao, Chin. Phys. C 45 (2021) 025109.
\bibitem{mazh14}
S. H. Mazharimousavi, M. Halilsoy, O. Gurtug, Eur. Phys. J. C 74 (2014) 2735.
\bibitem{has07}
M. Hassa{\"i}ne, C. Mart{\'i}nez, Phys. Rev. D 75 (2007) 027502.
\bibitem{has08}
M. Hassa{\"i}ne, C. Mart{\'i}nez, Class. Quantum Grav. 25 (2008) 195023.
\bibitem{hen10}
S. H. Hendi, Prog. Theor. Phys. 124 (2010) 493.
\bibitem{pan19}
G. Panotopoulos, {\'A}. Rinc{\'o}n, Int. J. Mod. Phys. D 28 (2019) 1950016.
\bibitem{tat20}
M. B. Tataryn, M. M. Stetsko, Int. J. Mod. Phys. D 29 (2020) 2050111.
\bibitem{gon09}
H. A. Gonzalez, M. Hassa{\"i}ne, C. Martinez, Phys. Rev. D 80 (2009) 104008.
\bibitem{born34}
M. Born, L. Infeld, Proc. Roy. Soc. Lond. A 144 (1934) 425.
\bibitem{fer03}
S. Fernando, D. Krug, Gen. Relativ. Gravit. 35 (2003) 129.
\bibitem{ban12}
R. Banerjee, D. Roychowdhury, Phys. Rev. D 85 (2012) 044040.
\bibitem{ban12_2}
R. Banerjee, D. Roychowdhury, Phys. Rev. D 85 (2012) 104043.
\bibitem{li16}
S. Li, H. L{\"u}, H. Wei, J. High Energ. Phys. 07 (2016) 004.
\bibitem{myun08}
Y. S. Myung, Yong-Wan Kim, Young-Jai Park, Phys. Rev. D 78 (2008) 084002.
\bibitem{dey04}
T. K. Dey, Phys. Lett. B 595 (2004) 484.
\bibitem{cai04}
Rong-Gen Cai, Da-Wei Pang, A. Wang, Phys. Rev. D 70 (2004) 124034.
\bibitem{mis08}
O. Mi{\v s}kovi{\'c}, R. Olea, Phys Rev. D 77 (2008) 124048.
\bibitem{wei10}
Yi-Huan Wei, Chin. Phys. B 19 (2010) 090404.
\bibitem{gunas12}
S. Gunasekaran, D. Kubiz{\v n}{\'a}k, R. B. Mann, J. High Energ. Phys. 11 (2012) 110.
\bibitem{zou14}
De-Cheng Zou, Shao-Jun Zhang, B. Wang, Phys. Rev. D 89 (2014) 044002.
\bibitem{bret17}
N. Bret{\'o}n, T. Clark, S. Fernando, Int. J. Mod. Phys. D 26 (2017) 1750112.
\bibitem{hen_all_14}
S. H. Hendi, M. Allahverdizadeh, Adv. High Energy Phys. (2014) 390101.
\bibitem{shey14}
A. Sheykhi, S. Hajkhalili, Phys. Rev. D 89 (2014) 104019.
\bibitem{hen16}
S. H. Hendi, S. Panahiyan, B. E. Panah, Int. J. Mod. Phys. D 25 (2016) 1650010.
\bibitem{tat19}
M. B. Tataryn, M. M. Stetsko, Int. J. Mod. Phys. D 28 (2019) 1950160.
\bibitem{jimen18}
J. B. Jim{\'e}nez, L. Heisenberg, G. J. Olmo, D. Rubiera-Garcia, Phys. Rep. 727 (2018) 1.
\bibitem{gaet17}
P. Gaete, J. A. Helay{\"e}l-Neto, Eur. Phys. Lett. 119 (2017) 51001.
\bibitem{krugl15}
S. I. Kruglov, Int. J. Geom. Methods Mod. Phys. 12 (2015) 1550073.
\bibitem{shap91}
A. D. Shapere, S. Trivedi, F. Wilczek, Mod. Phys. Lett. A 6 (1991) 2677.
\bibitem{sen93}
A. Sen, Nucl. Phys. B 404 (1993) 109.
\bibitem{cham20}
A. H. Chamseddine, W. A. Sabra, Phys. Lett. B 485 (2000) 301.
\bibitem{hart07}
S. A. Hartnoll, P. K. Kovtun, Phys. Rev. D 76 (2007) 066001.
\bibitem{hart07_2}
S. A. Hartnoll, P. K. Kovtun, M. Muller, S. Sachdev, Phys. Rev. B 76 (2007) 144502.
\bibitem{alba08}
T. Albash, C. V. Johnson, J. High Energ. Phys. 09 (2008) 121.
\bibitem{wirs01}
M. Wirschins, A. Sood, J. Kunz, Phys. Rev. D 63 (2001) 084002.
\bibitem{lu13}
H. L{\"u}, Y. Pang, C. N. Pope, J. High Energ. Phys. 11 (2013) 033.
\bibitem{dut13}
S. Dutta, A. Jain, R. Soni, J. High Energ. Phys. 12 (2013) 060.
\bibitem{eiroa06}
E. F. Eiroa, Phys. Rev. D 73 (2006) 043002.
\bibitem{gib95}
G. W. Gibbons, D. A. Rasheed, Nucl. Phys. B 454 (1995) 185.
\bibitem{stef07}
I. Zh. Stefanov, S. S. Yazadjiev, M. D. Todorov, Phys. Rev. D 75 (2007) 084036.
\bibitem{chem08}
W. A. Chemissany, Mees de Roo, S. Panda, Class. Quantum Grav. 25 (2008) 225009.
\bibitem{krug17}
S. I. Kruglov, Ann. Phys. 383 (2017) 550.
\bibitem{krug19}
S. I. Kruglov, Gen. Relativ. Gravit. 51 (2019) 121.
\bibitem{meng21}
K. Meng, L. Cao, J. Zhao, T. Zhou, F. Qin, M. Deng, Phys. Lett. B 819 (2021) 136420.
\end{thebibliography}
\end{document}